\documentclass[twocolumn]{aastex631}

\usepackage{graphicx}
\usepackage[flushleft]{threeparttable}
\usepackage{rotating}
\usepackage{amsmath,bm}
\received{December 22, 2023}
\accepted{February 9, 2024}

\begin{document}

\title{Classification of Enhanced Geoeffectiveness Resulting from High-Speed Solar Wind Streams Compressing Slower Interplanetary Coronal Mass Ejections }

\author[0000-0002-2655-2108]{Stephan~G.~Heinemann}
\affiliation{Department of Physics, University of Helsinki, P.O. Box 64, 00014, Helsinki, Finland}

\author[0000-0003-4236-768X]{Chaitanya~Sishtla}
\affiliation{Department of Physics, University of Helsinki, P.O. Box 64, 00014, Helsinki, Finland}

\author[0000-0002-4921-4208]{Simon~Good}
\affiliation{Department of Physics, University of Helsinki, P.O. Box 64, 00014, Helsinki, Finland}

\author[0000-0002-6373-9756]{Maxime~Grandin}
\affiliation{Department of Physics, University of Helsinki, P.O. Box 64, 00014, Helsinki, Finland}

\author[0000-0003-1175-7124]{Jens~Pomoell}
\affiliation{Department of Physics, University of Helsinki, P.O. Box 64, 00014, Helsinki, Finland}



\begin{abstract}
High-speed solar wind streams (HSSs) interact with the preceding ambient solar wind to form Stream Interaction Regions (SIRs), which are a primary source of recurrent geomagnetic storms. However, HSSs may also encounter and subsequently interact with Interplanetary Coronal Mass Ejections (ICMEs). In particular, the impact of the interaction between slower ICMEs and faster HSSs, represents an unexplored area that requires further in-depth investigation. This specific interaction can give rise to unexpected geomagnetic storm signatures, diverging from the conventional expectations of individual SIR events sharing similar HSS properties. Our study presents a comprehensive analysis of solar wind data spanning from 1996 to 2020, capturing 23 instances where such encounters led to geomagnetic storms ($SymH$ $< -30$ nT). We determined that interaction events between preceding slower ICMEs and faster HSSs possess the potential to induce substantial storm activity, statistically nearly doubling the geoeffective impact in comparison to SIR storm events. The increase in the amplitude of the $SymH$ index appears to result from heightened dynamic pressure, often coupled with the concurrent amplification of the CMEs rearward $|B|$ and/or $B_z$ components.

\end{abstract}



\section{Introduction} \label{sect:into}
High-speed solar wind streams (HSSs) were first observed by \cite{Snyder1963}, unveiling a link between recurring geomagnetic activity and the velocity of the solar wind recorded by Mariner 2. These HSSs were subsequently associated with dark regions on the Sun known as solar coronal holes \citep[][]{1967newkirk, wilcox68}{}{}. HSSs interact with the preceding slow solar wind to form Stream-- and subsequently, Corotating Interaction Regions (SIRs/CIRs) which gives rise to the creation of shocks, compression and rarefaction regions as well as forward and reverse waves \citep[][]{Belcher1971}{}{}. These are widely recognized as the origin of recurrent geomagnetic activity at approximately 27-day intervals, affecting various conditions in the Earth's atmosphere, ionosphere, and magnetosphere \citep[\textit{e.g.,}][]{alves06,Grandin2015,Grandin2017,kilpua17,2018richardson}. From in-situ measurements at 1au, \cite{1990schwenn} defined HSSs as solar wind with speeds of approximately $v_p \approx 400$ to $800 \, \text{km/s}$, low densities $(n_p \approx 3 \, \text{cm}^{-3})$, and high temperatures $(T_p \approx 2.5 \times 10^5~\mathrm{K})$. \\

In contrast to the continuous and quasi-periodic outflow of the solar wind, Coronal Mass Ejections (CMEs) and their in-situ measured counterpart, Interplanetary Coronal Mass Ejections (ICMEs) have the capacity to abruptly disrupt the solar \textit{and interplanetary} structure and are typically the primary drivers of intense geomagnetic disturbances \citep[\textit{e.g.,}][]{1997farrugia,schwenn06, 2010richardson_RC-list, Temmer2021LRSP}{}{}. These effects may be enhanced due to an interaction with other magnetic structures propagating in the heliosphere, primarily due to the compression of the existing $B_z$ magnetic field component \citep[\textit{e.g.,}][]{Farrugia2006,Dumbovic2015,2017lugaz}. The examination of interaction events involving CMEs \citep[\textit{e.g.,} see][]{Kilpua2019,Scolini2020}{}{} and their resulting geoeffectiveness has been recognized as one of the objectives outlined by the heliospheric clusters within the International Space Weather Action Teams (ISWAT), which serves as a roadmap for future research \citep[][]{TEMMER2023}{}{}.\\

The interaction between CMEs and the solar wind, particularly HSSs, can lead to significant alterations in CME properties during their journey through the heliosphere. These changes encompass potential deformations, kinks, rotations, and erosions of the embedded flux rope, along with deflections and the amplification of turbulence within the sheath region \citep[\textit{e.g.,}][]{2004manchester,2013Isavnin,2014lavraud,2015lugaz,kilpua17,2019zhuang,2019Heinemann_hsscme}{}{}. This could result in alterations to the anticipated geoeffectiveness of the transient. However, in instances where the interaction takes place in closer proximity to the Sun, the typical cause of geoeffectiveness is usually attributed to the ICME itself.
\\

In this study, we explore the geomagnetic impact of a specific type of ICME-HSS events in comparison to typical slow-fast wind interaction events (SIRs). It entails a statistical analysis of the potential amplification of geomagnetic effects resulting from HSSs encountering slow preceding ICMEs that act as magnetic obstacles. We find that this particular category of solar wind interaction can substantially intensify the geomagnetic storm caused by the HSS in contrast to a typical SIR. In Section~\ref{sect:data}, we present the data followed by the analysis results in Section~\ref{sect:res}. We discuss and summarize the findings in Section~\ref{sect:disc}.

\section{Dataset} \label{sect:data}

In this study we use solar wind plasma and magnetic field in-situ measurements provided by the OMNI database \citep[][]{papitashvili2020omni_1h,papitashvili2020omni5min}{}{}. This entails intercalibrated data from different spacecraft (primarily from the \textit{Advanced Composition Explorer} \citep[ACE;][]{1998stone_ACE}{}{} and \textit{Wind} \citep{1995acuna_GSS}) that is propagated to the Earth's bow-shock nose.\\

Through manual classification of plasma and magnetic field data spanning from 1996 to 2020, we identify a specific category of geoeffective HSS-ICME interaction events, where a slower preceding ICME is pushed by a subsequent faster HSS. It is classified based on the criterion that the average speed of the ICME is slower than the peak velocity of the HSS based on in-situ measurements near 1au. We argue that in these geomagnetic storm events, the main driver of the heightened geoeffectiveness is the HSS, with the source likely stemming from a combined effect of both the HSS and the ICME due to their interaction. We define a geomagnetic storm as a time period during which the magnetic perturbation associated with the ring current \citep[measured by the $SymH$ index][]{Wanliss2006} falls below $-30$~nT \citep[following][]{Loewe1997}. In practice, we focus on events where the minimum $SymH$ coincides temporally with the HSS-CME interaction region. Our definition of the interaction is twofold. An interaction between a HSS and a preceding slower ICME is identified if there is no undisturbed ambient solar wind between those transients. The interaction region encompasses both the compressed and/or disturbed rear part of the ICME and the compressed boundary layer between the ICME and the undisturbed HSS. We omitted all events where the geoeffectiveness can be clearly linked to the impact of solely the ICME or the fast wind. Applying these criteria, we identified 23 such events over the 24-year period, spanning more than two solar cycles.  \\

To examine the in-situ properties of the interacting structures, we classified the boundaries of the magnetic ejecta using the criteria following \cite{Klein1982}, \cite{Lepping1990} and \cite{2010richardson_RC-list}. The boundaries for the compressed boundary layer (BL) and the start of the undisturbed HSS (referred to as HSS) were determined according to the criteria outlined by \cite{Belcher1971}. We define the boundary layer as a potential mix of (i) solar wind compressed by the trailing HSS, (ii) some wind that previously formed part of the HSS that has slowed down due to the interaction, and, in cases where the trailing ICME boundary is poorly defined, (iii) some ICME material.  It does not include compressed regions that clearly fall inside the ICME rear. We intentionally refrain from defining the ICME's sheath region and the end of the HSS interval in this study since our primary focus is solely on the interaction itself. In this study, we further refer to the in-situ measured magnetic ejecta containing a flux rope-like structure as the ICME. In Appendix~\ref{app:supp}, Tables~\ref{tab:list} and~\ref{tab:props} respectively list the calculated parameters and present the associated event properties. These include the derived values for the timing of the boundaries as defined above, minimum $SymH$ values associated with the interaction, average CME velocity, peak velocity of the HSS, maximum Interplanetary Magnetic Field (IMF) magnitudes and minimum $B_z$ values in different sections, variation of the magnetic field magnitude in the boundary layer, and peak plasma density and pressure. Additionally, we determined the configuration of the magnetic flux rope and helicity by analyzing the respective magnetic field hodograms according to the criteria given by \cite{Zhao1996}, \cite{Bothmer1997}, \cite{1998bothmer+schwenn} and \cite{Mulligan1998}.

\section{Results} \label{sect:res}

\begin{figure}
\centering \includegraphics[width=1\linewidth,angle=0]{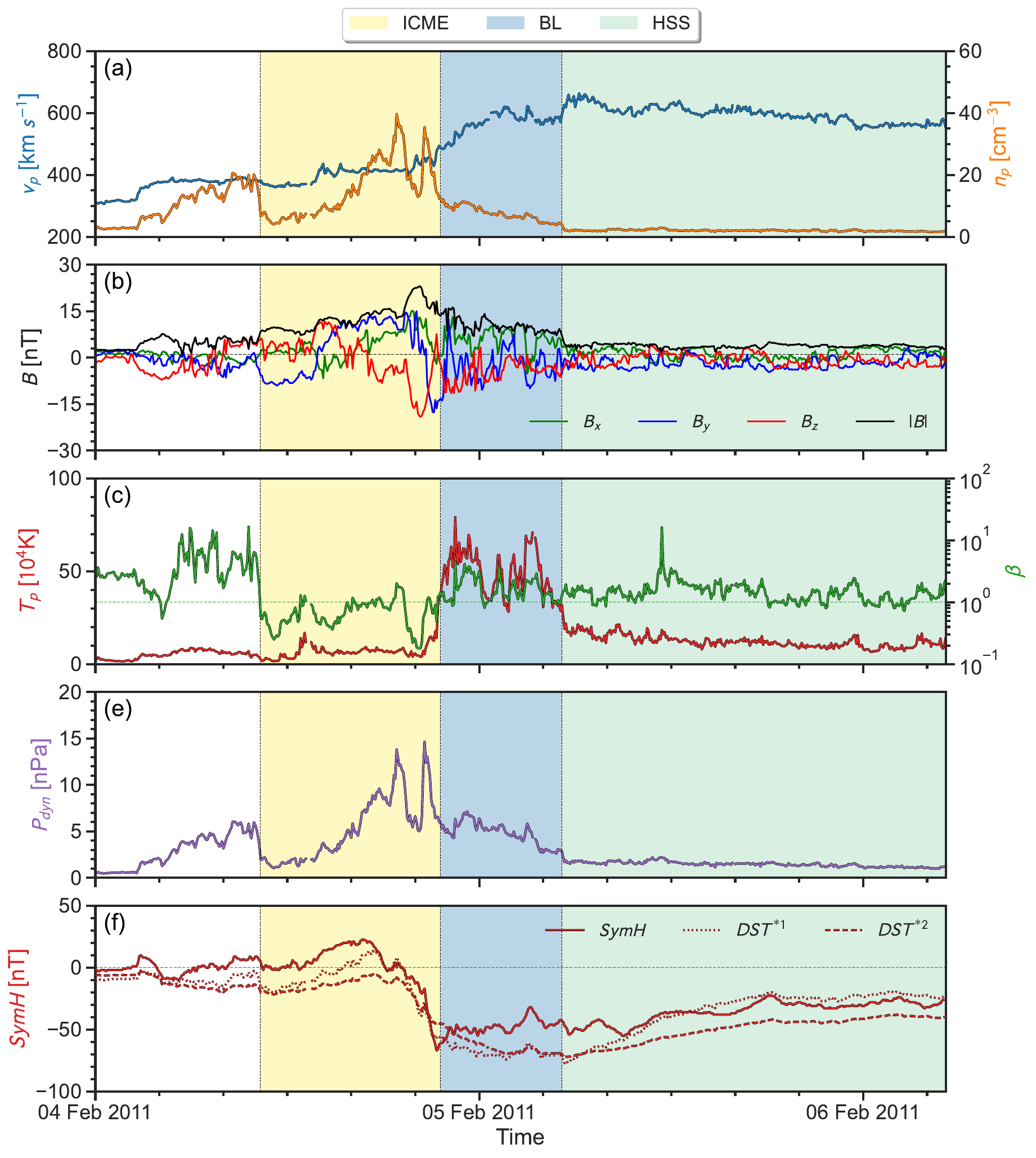}
\caption{Example of a HSS interacting with a slower ICME leading to an intensified geomagnetic response in February 2011. Panel (a) displays the solar wind velocity (in blue) and number density (in orange), while panel (b) presents the total magnetic field strength and its components in GSM coordinates. Panel (c) exhibits proton temperature (in red) and plasma beta (in green), followed by (d) the dynamic solar wind pressure and (f) the geomagnetic activity index, $SymH$ (in red). The dashed and dotted lines in panel (f) correspond to the predicted DSTs according to Appendix}~\ref{app:dst_pred}. The vertical sections mark the boundaries of the magnetic structure, boundary layer, and HSS.\label{fig:overview}
\end{figure}

In Fig.~\ref{fig:overview}, we present an event from February 2011, which serves as an illustrative example of the observed structure resulting from the interaction of various solar wind domains. We observe a \textit{ICME} structure propagating at a speed of $\sim 400$~km~s$^{-1}$ slightly higher than that of the slow solar wind. This is followed by the significantly faster HSS plasma. Notably, the magnetic field exhibits a distinct rotation, suggestive of a flux rope structure. For our events, we predominantly find a North-South configuration of the $B_z$ component, as exemplified in this case and highlighted in Fig.~\ref{fig:FR_orientation} in  Appendix~\ref{app:supp}. Towards the rear of the magnetic structure, there is an abrupt change in both the total magnetic field strength ($|B|$) and its $B_z$ component aligning with a general density enhancement. We believe the magnetic field and density increase is linked to the interaction between the slower ICME and the faster HSS. It looks like the faster HSS is compressing the rear of the slower ICME, causing the magnetic field enhancement which we consistently observe  across our events. Following the magnetic structure, we encounter a zone with fluctuating magnetic fields, higher proton temperatures, and increased velocities. The values of density, temperature, and magnetic field in this region all exceed those of the subsequent HSS segment, which we identify as the undisturbed HSS. The geomagnetic response temporally coincides with the compressed rear section of the magnetic structure and/or this boundary layer. To assess the cause of the geomagnetic disturbance and subsequently estimate its strength using an analytical model, we utilize the DST (Disturbance Storm-Time Index, in principle a low resolution version of the $SymH$ index) prediction method initially proposed by \cite{Burton1975}. Here we use the applications derived by \cite{Fenrich1998} (further called $DST^{*1}$) and \cite{OBrien2000} ($DST^{*2}$). Notes on how the DST is estimated are shown in Section~\ref{app:dst_pred}. For our 23 events we find the calculated DST to be matching the observed $SymH$ index very well (with average \textit{Pearson} correlation coefficients of $0.87$ and $0.81$ for $DST^{*1}$ and $DST^{*2}$ respectively).\\

We correlate the derived parameters (listed in Tab.~\ref{tab:props} in Appendix~\ref{app:supp}) using the \textit{Spearman} correlation coefficient. A correlation matrix of all properties is shown in Fig.~\ref{fig:corr_matrix} in Appendix~\ref{app:supp}. Note that we have excluded one event from the correlation analysis as an outlier due to the strong $SymH$ ($-218$~nT) and $-B_z$ ($-55$~nT) values. This extreme case skews the correlation too strongly. From our analysis we find that $SymH$ is not significantly correlated with any properties of the CME, HSS or the boundary layer. We find a weak and not significant correlation with the maximum values of the ICMEs magnetic field ($|B|$ and $|B_z|$), which corresponds to the compressed rear part of the ICME. We find that the properties of the boundary layer are primarily correlated with the speed difference between the CME and HSS, $\Delta \mathbf{v}_{HSS-CME}$. A larger difference leads to a stronger compression, resulting in higher magnetic field, magnetic field fluctuation, density and pressure within the boundary layer. However, it seems that this is not directly correlated to the $SymH$ index. \\

The profiles of all other identified events are similar to the one presented in Fig.~\ref{fig:overview}, although some differences are present. The majority of the events feature a flux rope that is oriented in the ecliptic ($78\%$) with North--South rotation in $B_z$ ($61\%$), and only a few flux ropes seem to be oriented out of the ecliptic ($17\%$). For one event the flux rope orientation could not be determined. Events associated with a North-West-South (NWS) configuration flux ropes statistically appear to have the strongest geomagnetic impact, although the limited number of occurrences prevents making a conclusive statement. In addition, we find that the helicity of the flux rope (right-handed vs. left-handed) does not affect the geoeffectivness of this specific category of interaction events (11 right-handed and 12 left-handed flux ropes were found). We visualize this in Fig.~\ref{fig:FR_orientation} in Appendix~\ref{app:supp}. We further observe clear indications that $74\%$ of events exhibit a FR structure that is compressed at the rear. This is evidenced by an enhanced magnetic field, averaging a $25\%$ increase, in the rear part of the FR. Specifically, the mean of $|B|$ in the last $30\%$ of the FR exceeds the mean $|B|$ of the first $70\%$ of the FR. This emphasizes the importance of the presence of a magnetic structure in enhancing geomagnetic activity in these interaction events. \\

\begin{figure}
\centering \includegraphics[width=1\linewidth,angle=0]{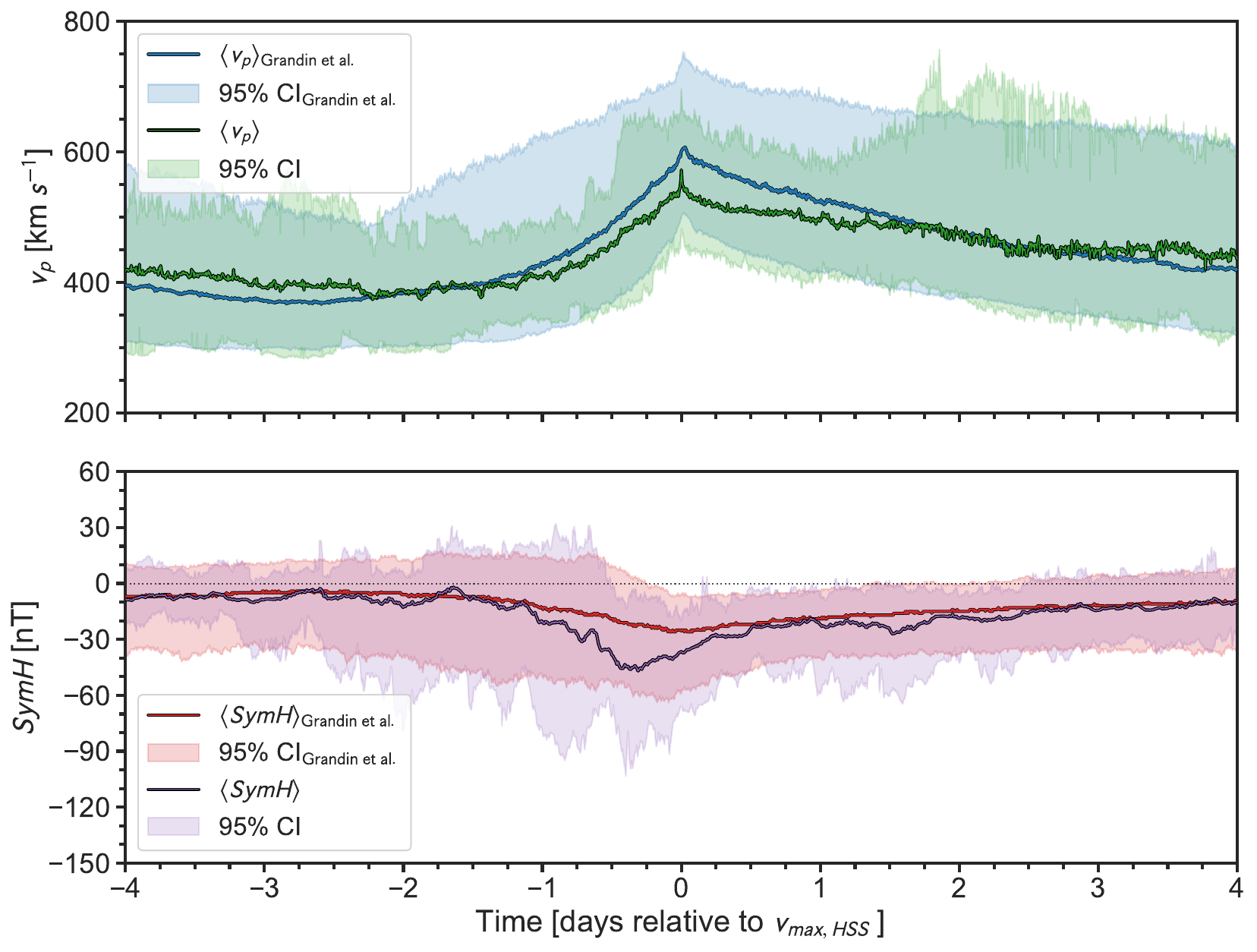}
\caption{Superposed epoch analysis of solar wind velocity ($v_p$; top panel) and the geomagnetic activity index ($SymH$, bottom panel) centered on the time when the HSS velocity reaches its maximum. The colored lines and shaded areas (velocity in blue, geomagnetic index in red) denoted with Grandin et al. represent the median and the 95th percentiles of the 478 SIRs with $SymH_{\mathrm{min}} < -30~\mathrm{nT}$ detailed in \cite{Grandin2019}. Overlaid in green and purple are the results from our 23 events.}\label{fig:superposed}
\end{figure}

In Fig.~\ref{fig:superposed}, we performed a superposed epoch analysis using the 478 of the 642 SIRs documented in \cite{Grandin2019}. The selected SIRs all led to a geomagnetic storm ($SymH < -30$~nT), employing the same criteria as for our events. Therefore, we can analyze the distribution within the same parameter regime. We perform the same analysis for our events.
We observe that the distribution of velocities in our events is average compared to the set of 478 SIR events. The median peak velocity (Fig.~\ref{fig:superposed}, top panel) for our events ($572$~km~s$^{-1}$) is $35$~km~s$^{-1}$ lower than for the set by Grandin et al. ($607$~km~s$^{-1}$). It is revealed that the peak in geoeffectiveness (minimum in $SymH$; Fig.~\ref{fig:superposed}, bottom panel) for SIRs is observed to occur in temporal alignment with the peak in HSS velocity. In this study, for this specific set of ICME-HSS interaction events, the peak in geoeffectiveness takes place, on average, $7.5$ hours earlier. This can be explained if the geoeffectiveness is linked to the interaction region between the two transients. The minimum in the $SymH$ index of the median curve in the superposed epoch analysis indicates a value of $-47$~nT ($-42$~nT if the strong outlier event is excluded), in contrast to the $-26$~nT (with a lower 95th percentile of $-64$nT) of the SIR events from Grandin et al.. This implies that the median ICME-HSS interaction event, as defined in this study, generates a geomagnetic effect equivalent to the 85th percentile of geomagnetic storms caused by SIRs. We acknowledge that the locations of the minimum DST of the SIR events may not always align with the location of maximum velocity. Consequently, when employing superposed epoch analysis, the $95\%$ confidence interval distribution covers a wide range. Thus leading to the fact that the minimum DST of the median curve ($-26$~nT) exceeds the selection threshold of $-30$~nT. This analysis underscores the substantial increase in geoeffectiveness resulting from the compound interaction between ICMEs and HSSs, statistically surpassing that of SIR events. \\

\begin{figure}
\centering \includegraphics[width=1\linewidth,angle=0]{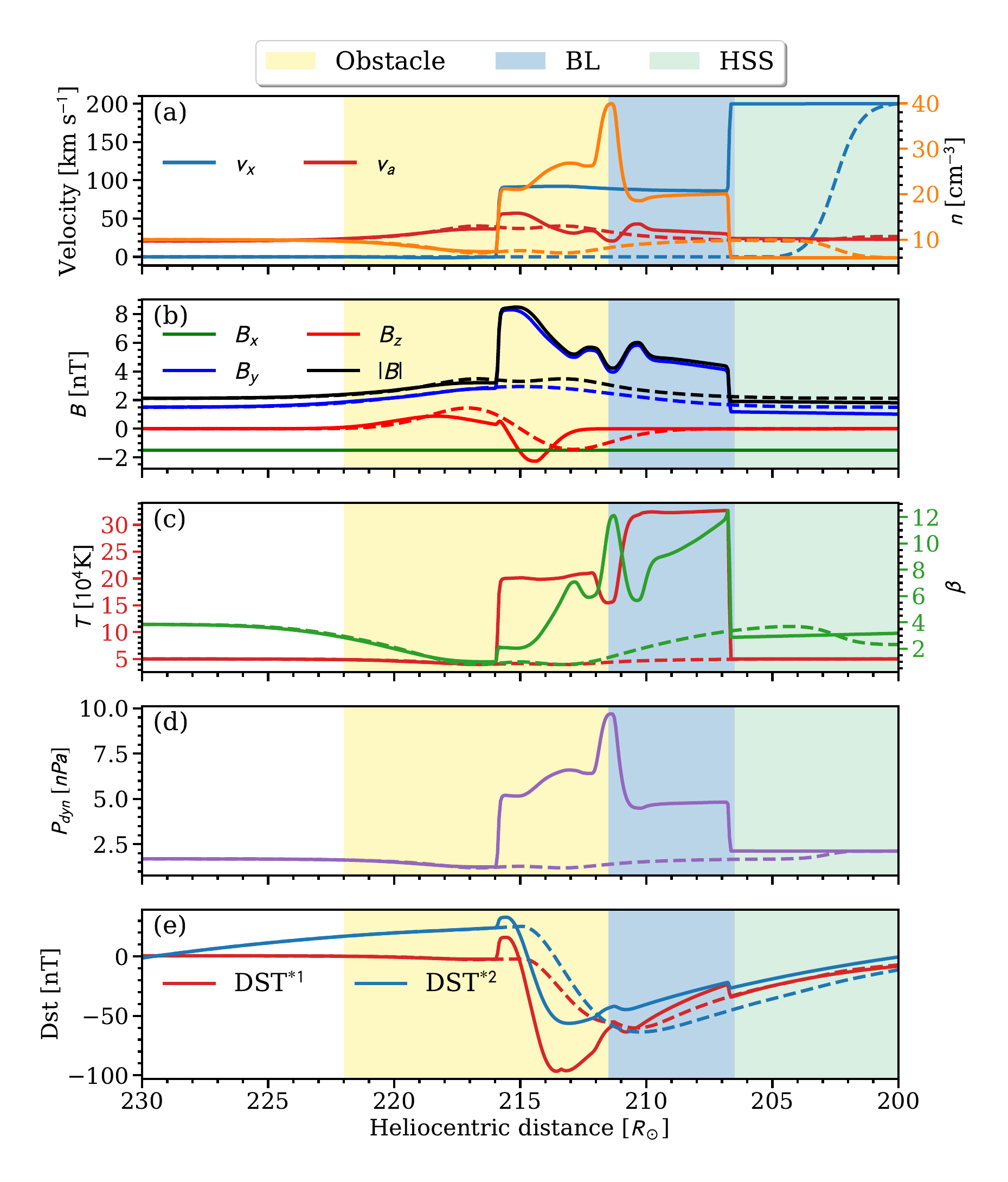}
\caption{MHD simulation of a HSS interacting with a magnetic obstacle in cartesian coordinates co-moving with the magnetic obstacle. The dashed lines indicate the initial condition of the simulation while solid lines present the solar wind plasma at $\approx 19.4$ hr simulation time. Panel (a) shows the solar wind $v_x$ velocity component (in blue), the Alfv\'en velocity $v_a = B/\sqrt{\mu_0\rho}$ (in red), and the number density $n$ (in green). The magnetic fields are presented in panel (b). Panel (c) presents the plasma temperature $T$ (in red), and the plasma beta $\beta$ (in purple). Finally, in panel (d) we show the simulated geomagnetic activity index DST (in red) according to~\ref{appsub:dst_sim}. The shaded regions mark the solar wind structures corresponding to the magnetic obstacle, HSS, and the compressed boundary layer at $\approx 19.4$ hr simulation time.}
\label{fig:sim}
\end{figure}

To demonstrate that the in-situ measured solar wind plasma and magnetic field profiles correspond to a HSS-CME interaction event, we employ a 1D magnetohydrodynamic (MHD) model that emulates an encounter between an HSS and a flux-rope-like magnetic obstacle (described in Appendix~\ref{app:model}). The model describes the interaction between a low-density, high-velocity plasma (HSS) and a magnetic obstacle containing a smooth rotation of the magnetic field (flux-rope) between $200$ solar radii ($R_\odot$) and $230~R_\odot$. The simulation (Fig.~\ref{fig:sim}) is performed in the frame of reference of the magnetic obstacle that is assumed to be co-propagating with the background solar wind at a uniform velocity. The HSS is introduced in the simulation via a time-dependent boundary condition at the $200~R_\odot$ boundary. The initial solar wind plasma is depicted by the dashed lines in Fig.~\ref{fig:sim}. The smooth transition from the quiet solar wind to the HSS is observed by the dashed sigmoidal $v_x$ profile (Fig.~\ref{fig:sim}a) which transitions from $v_x = 0$ (quiet solar wind) to $v_x = 200$ km s$^{-1}$ (HSS). A similar transition is applied for the number density from $n = 10$ cm$^{-3}$ (quiet solar wind) to $n = 3$ cm$^{-3}$ (HSS). The quiet solar wind and magnetic obstacle are at rest ($v_x = 0$), with the magnetic obstacle characterized by a smooth rotation in $B_z$, and an enhanced $B_y$. The magnetic obstacle has a lower density than the quiet wind to maintain an equilibrium with the surrounding solar wind (Equations~\ref{eq:pressure-balance} and~\ref{eq:entropy-balance}). The choice of modeling the solar wind in the frame of reference of the magnetic obstacle allows us to extend the simulation results to any constant solar wind flow speeds. Thus, by assuming a constant solar wind flow speed of $v_0 = 450$ km s$^{-1}$ in the $x$-direction we can estimate the dynamic pressure in Fig.~\ref{fig:sim}d. Finally, by further scaling the background magnetic fields to match solar wind observations (Appendix~\ref{appsub:dst_sim}), we plot the geomagnetic activity index Dst in Fig.~\ref{fig:sim}e. The $DST^{*1}$ and $DST^{*2}$ models for calculating the Dst index differ in the relative contributions of the convective electric field and dynamic pressure to the enhancement of the ring current. The calculation of Dst in the simulation is detailed in \ref{appsub:dst_sim}. The interaction of the HSS with the magnetic obstacle at $19.4$ hr simulation time (shown via solid lines in Fig.~\ref{fig:sim}) compresses the rear of the obstacle as well as the front of the HSS resulting in a density increase and an enhanced of the magnetic field, specifically the $B_z$ component. This results in a substantial increase in the estimated dynamic pressure leading to an enhancement in the Dst that can be linked to the interaction region. The similarity in the $DST^{*1}$ and $DST^{*2}$ models in Fig.~\ref{fig:sim}e validates the Dst calculation methodology employed for the simulation. We observe that the model results qualitatively replicate the observed solar wind profiles, as exemplified in Fig.~\ref{fig:overview}. Due to the higher speed of the HSS, it compresses the rear part of the magnetic obstacle, leading to an enhancement in the magnetic field, possibly in the $B_z$ component, as well as an increase in density. Additionally, it causes the solar wind to accumulate between the magnetic structure and the undisturbed high-speed wind, thereby creating a compression region referred to as the boundary layer in this study.

\section{Discussion and Conclusion} \label{sect:disc}

In our study, we highlighted a specific type of interaction between HSSs and slower ICMEs acting as magnetic barriers. This unique encounter can trigger enhanced geomagnetic storm signatures, a result not typically expected from similar SIR events. Analyzing solar wind data spanning from 1996 to 2020, we identified 23 events where 1) signatures of interaction between preceding slower ICMEs and successive faster HSSs were present, and 2) the enhanced geoeffectiveness could be attributed to the interaction between these two transients.\\

We observed that these interaction events consistently exhibit similar in-situ profiles across various parameters. This involves a preceding slow-moving magnetic structure traveling at or just slightly above the ambient solar wind speed, consisting of an ``undisturbed'' front segment and a compressed rear section. We define the ``undisturbed'' front segment as having experienced no interaction with the subsequent HSS. The compressed rear portion displays a marked increase in both magnetic field and density. Subsequently, a period of compressed boundary layer follows, characterized by fluctuating magnetic fields and significantly higher densities compared to the subsequent undisturbed fast-moving solar wind. The minimum observed in the $SymH$ index aligns with the segments of interaction, specifically the compressed regions of the ICME and the boundary layer. \cite{Sarkar2023}, who analyzed such an interaction event using STEREO and MESSENGER data, discovered a very similar profile in the in-situ measured data of STEREO. However, MESSENGER data did not show any interaction, suggesting that this event occurred further out in the heliosphere. Their conclusion was that, in addition to the enhanced dynamic pressure due to the compression, the CME was prevented from expanding in a self-similar manner, resulting in an overall increase in the flux rope magnetic field.\\

We find that the geoeffectiveness of these interaction events can be explained through the interaction between a HSS and a magnetic obstacle as shown in a 1D MHD model. The model results closely match the observed solar wind profiles generating sections comparable to the actual observations, demonstrating an enhancement in both magnetic field and density within the interaction regions. The calculated Dst indices imply that the increase in dynamic pressure due to the compression of magnetic field and plasma during the interaction of these two interplanetary phenomena may lead to an enhanced geomagnetic response. The compression of the magnetic obstacle by the HSS is a consequence of the HSS steepening into a shock and compressing the plasma ahead of it.\\

The HSSs examined in this study demonstrate fairly typical characteristics when compared to the statistical analysis conducted by \cite{Grandin2019}. Nonetheless, their geomagnetic response, as indicated by the $SymH$ index, was exceptional, positioning the median geoeffectiveness at the 85th percentile when compared to all SIRs that produced geomagnetic storms. The timing of the geomagnetic response, contrary to statistical expectations \citep[][]{Grandin2019}{}{}, does not align with the peak of the bulk velocity, but is instead, on average, $7.5$ hours earlier. It roughly correlates with the arrival of the boundary layer formed between the compressed rear of the ICME and the leading edge of the undisturbed HSS. ICMEs can typically exhibit strong geoeffectiveness; however, slow ICMEs with a relatively feeble magnetic field, particularly those with a weak $B_z$ component, tend to have notably reduced geoeffectiveness. Despite not observing a significant direct correlation, we believe that the presence of a magnetic structure, whose magnetic field --specifically the $Bz$ and $|B|$ components -- is enhanced due to compression by the HSS, plays a substantial role in the increase of geomagnetic activity. This emphasizes the significance of flux rope-like structures in magnifying geomagnetic effects. Overall, our results suggest that the geoeffectiveness of this specific ICME-HSS interactions is statistically stronger than HSS-slow wind interactions. In events where the $SymH$ minimum occurs near (within or just after) the ICME rear, we can infer that the HSS enhanced the geoeffectiveness of the ICME. Conversely, when the minimum occurs at the time farther away from the ICME rear end, we can infer that the ICME enhanced the geoeffectiveness of the HSS. \\

Studies such as \cite{Verbanac2013} have identified an inverse relationship between the DST index and ICME speed (Pearson's correlation coefficient: $cc_{\mathrm{Pearson}}=-0.57$) as well as CME peak magnetic field (Pearson's correlation coefficient: $cc_{\mathrm{Pearson}}=-0.85$). Similar findings were corroborated by \cite{Shen2017} ($cc_{\mathrm{Pearson}}=-0.77$) in their examination of the correlation between DST and magnetic field strength. In our study, where the magnetic field of the magnetic ejecta is consistently below $20$~nT, we find only a very weak, not significant correlation of $|B|$ to the $SymH$ index ($cc_{\mathrm{Spearman}}=-0.35$) and no correlation between the $SymH$ index and the peak velocity of the HSS ($cc_{\mathrm{Spearman}}=0.14$) or average ICME velocity ($cc_{\mathrm{Spearman}}=0.24$). While we do not observe a correlation between the properties of the boundary layer and the $SymH$ index, it does exhibit a correlation with the speed difference between the ICME and HSS (Please refer to Fig.~\ref{fig:corr_matrix} in Appendix~\ref{app:supp} for a comprehensive overview of all correlations). These results are not entirely unexpected as the solar wind -- magnetosphere coupling is a complex topic that has yet to be fully solved. \\

From statistically analyzing and describing a particular type of ICME-HSS encounter, we can derive the following conclusions:
\begin{itemize}
    \item In these specific ICME-HSS interactions, we observe that the geoeffectiveness is statistically stronger than in slow wind-HSS interactions (\textit{i.e.,} SIRs). The median geoeffectiveness of the events studied is positioned at the 85th percentile in comparison to all SIRs that induced geomagnetic storms.
     \item While the enhanced geoeffectivity is a compound effect of the interacting ICME and HSS, a $SymH$ minimum occurring close to the ICME rear may be interpreted as being due to an enhancement of the ICME geoeffectiveness; $SymH$ minima occurring later in the interaction region may likewise be attributed to an enhancement of the HSS geoeffectiveness.   
  
    \item The increase in the amplitude of the $SymH$ index is likely a combined effect of heightened dynamic pressure, along with the frequently observed amplification of the rearward $|B|$ and/or $B_z$ components in the preceding ICME.
    
\end{itemize}

While our results demonstrate that all the observed events resulted in geomagnetic storms, it is important to note that this was a selection parameter. Consequently, while we can conclusively demonstrate that the interaction significantly amplifies geoeffectiveness, we cannot draw conclusions regarding the frequency and overall prevalence.

\section*{Acknowledgments}
This research was funded by the Austrian Science Fund (FWF) Erwin-Schr\"odinger fellowship J-4560. Open access funded by Helsinki University Library. SGH and CS express their gratitude to the Space Physics Group in Helsinki for their unwavering support during a challenging period of emotional strain caused by our efforts in predicting the DST index. MG's, SG's, JP's, and CS's work is funded by the Research Council of Finland (grants 338629-AERGELC'H, 338486-, 346612-INERTUM, and 343581-SWATCH respectively).

%






\newpage
\appendix

\section{Empirical Dst prediction from solar wind data}\label{app:dst_pred}
The prediction of the Dst index from solar wind conditions is expressed as~\cite{Burton1975},
\begin{align}
    \frac{d \mathrm{Dst}^*}{dt} = Q(t) - \frac{\mathrm{Dst}^*}{\tau},
\label{eq:Dst-eq}
\end{align}
where $Q(t)$ is proportional to the rate of energy injection into the ring current, $\tau$ is the decay time, and Dst$^*$ is the corrected form of the Dst index with the magnetopause currents removed,
\begin{align}
    \mathrm{Dst}^* = \mathrm{Dst} - b\sqrt{\mathrm{P_{dyn}}} + c.
\label{eq:Dst_star}
\end{align}
Here $\mathrm{P_{dyn}}$ is the dynamic pressure, and $b,~c$ are parameters fitted from observational data. The specific choices of $Q(t),~\tau,~b,~c$ inform the impact of the dynamic pressure and convective electric field on the energy transfer to the magnetosphere. The general estimation of $Q(t)$ can be expressed in terms of $VBs$, the dawn-to-dusk component of the interplanetary electric field. This can be computed from in-situ measurements in Geocentric Solar Magnetospheric (GSM) coordinates using the following method:
\begin{align}
    VBs~  &=   \begin{cases}
    |VB_z|& B_z < 0 \\
    0 & B_z \geq 0
    \end{cases} &[\mathrm{mV/m}]  
\intertext{In this study, we use two specific models designated $DST^{*1}$~\citep{Fenrich1998} and $DST^{*2}$~\citep{OBrien2000}. The $DST^{*1}$ model is based on the assumption that a higher dynamic pressure facilitates more energy transfer to the atmosphere, while $DST^{*2}$ takes into account the movement of the outer boundary of the ring current to lower altitudes by an enhanced convective electric field. The differences between the $DST^{*1}$, $DST^{*2}$, and the original \citet{Burton1975} model are detailed in \citet{OBrien2000Forecasting}, along with the choices of the $Q(t),~\tau,~b,~c$ parameters. The $DST^{*1}$ model uses the following parameters:}
    Q~  &=   \begin{cases}
      ~0 & VBs \leq 0.5\\
      ~-4.32 (VBs -0.5) P_{dyn}^{1/3}   &  VBs > 0.5
    \end{cases} &[\mathrm{nT/h}]  \\
    \tau~ &=   \begin{cases}
      ~7.7 & VBs \leq 4\\
      ~3 &  VBs > 4
    \end{cases}  &[\mathrm{h}] \\
    b~ &= 15.8 &[\mathrm{nT/\sqrt{nPa}}]\\
    c~ &= 20 &[\mathrm{nT}]
\intertext{wheras the $DST^{*2}$ model uses:}
      Q~  &=~-4.4 \times   (VBs -0.5) &[\mathrm{nT/h}]\\
    \tau~ &=   2.4 \times exp\left[\frac{9.74}{4.69+VBs}\right] &[\mathrm{h}]\\
    b~ &= 7.26 &[\mathrm{nT/\sqrt{nPa}}] \\
    c~ &= 11 &[\mathrm{nT}]
\end{align}
Finally, Equation~\ref{eq:Dst-eq} can be solved as,
\begin{align}
    \mathrm{Dst}^*(t + \Delta t) = \mathrm{Dst}^*(t) + \left[Q(t) - \frac{\mathrm{Dst}^*(t)}{\tau}\right]\Delta t
\label{eq:dst-difference-eq}
\end{align}
where $\Delta t$ is the time cadence of the measuring instrument, the start time ($t_0$) is set to 12 hours before the start of the magnetic structure and the Dst at the start time was set to 0 ($Dst(t_0) = 0$).

\newpage
\section{MHD model for a HSS interacting with a magnetic obstacle}\label{app:model}
We perform a MHD numerical simulation assuming variations in one spatial dimension (along the $x$ direction) but retaining all three components of velocity and electromagnetic fields. The MHD equations are solved in Cartesian coordinates, with the magnetic field ensured to be divergence free to the floating point accuracy by utilizing the constrained transport method~\citep{Kissmann2012}. The relevant equations are
\begin{align}
    \frac{\partial \rho}{\partial t} + \nabla \cdot (\rho \mathbf{v}) = 0,
\label{eq:mass-cont}
\end{align}
\begin{align}
    \frac{\partial (\rho\mathbf{v})}{\partial t} + \nabla \cdot [\rho\mathbf{v}\mathbf{v} + (P+\frac{B^2}{2\mu_0})\mathbf{I} - \frac{\mathbf{B}\mathbf{B}}{\mu_0}] = 0,
\label{eq:mom-cont}
\end{align}
\begin{align}
    \frac{\partial \mathcal{E}}{\partial t} + \nabla\cdot[(\mathcal{E} + P - \frac{B^2}{2\mu_0})\mathbf{v} + \frac{1}{\mu_0}\mathbf{B}\times(\mathbf{v}\times\mathbf{B})] = 0,
\label{eq:energy-cont}
\end{align}
\begin{align}
    \frac{\partial \mathbf{B}}{\partial t} - \nabla \times (\mathbf{v}\times\mathbf{B}) = 0,
\label{eq:induction}
\end{align}
with
\begin{align}
    \mathcal{E} = \frac{1}{2}\rho v^2 + \frac{P}{\gamma-1} + \frac{B^2}{2\mu_0},
\label{eq:def:energy}
\end{align}
where $\rho$, $\mathbf{v}$, $\mathbf{B}$, $\mathcal{E}$, and $P$ correspond to the mass density, bulk plasma velocity, magnetic field, total energy density, and thermal pressure. The MHD equations (Equations~\ref{eq:mass-cont}-\ref{eq:induction}) are advanced in time using the strong stability preserving (SSP) Runge-Kutta method~\citep{Pomoell2012}, and employ the Harten–Lax–van Leer (HLL) approximate Riemann solver supplied by piece wise, linear slope-limited interface states. These methods have been used in previous studies of the solar corona~\citep{Sishtla2022, Sishtla2023}. The numerical simulation uses a polytropic index of $\gamma = 5/3$, and assumes a ideal-gas equation of state $P = (\rho/m)k_B T$ where $k_B$ is the Boltzmann's constant, $m = m_p/2$ is the mean molecular mass, and $m_p$ is the proton mass. The simulation domain extends from $200$ solar radii ($R_\odot$) to $230~R_\odot$, with the dynamical quantities linearly extrapolated at the $230~R_\odot$ boundary to enforce solar wind outflow. The boundary condition at $200~R_\odot$ is utilised to inject a HSS into the simulation.

\subsection{Simulation setup}
The simulation is initialised to model both the HSS and the magnetic obstacle. The magnetic obstacle is assumed to be embedded and co-propagating in the background quiet solar wind in the $x$-direction. We enforce MHD equilibrium on the magnetic obstacle and the surrounding wind to model steady rotations of the magnetic field components to mimic variations typical of flux ropes. Thus, operating in the frame of reference of the quiet solar wind ($v_x = 0$) and assuming a constant $B_x = B_0 \hat{x}$, we require (from Equations~\ref{eq:mass-cont}-\ref{eq:induction})
\begin{align}
    P + \frac{B^2}{2\mu_0} = \mathrm{constant}
\label{eq:pressure-balance}
\end{align}
\begin{align}
    P \rho^{-\gamma} = \mathrm{constant}
\label{eq:entropy-balance}
\end{align}
where $B = B_0\hat{x} + B_y(x) \hat{y} + B_z(x)\hat{z}$. We model the magnetic obstacle by choosing $B_z(x)$ to be the subtraction of two Gaussian functions with means located at $214~R_\odot$ and $216~R_\odot$, and widths (standard deviation) of $2~R_\odot$ each, and $B_y(x)$ to be a Gaussian located at $215~R_\odot$ and a width of $4~R_\odot$. Taking $B_0 = -1.5$ nT we can solve for the equilibrium $P(x)$ and $\rho(x)$ through Equations~\ref{eq:pressure-balance} and~\ref{eq:entropy-balance}. The HSS is modeled via an increased velocity $v_x$ and a decreased density near the $200~R_\odot$ boundary. We initialized the HSS by accounting for a smooth transition from the quiet solar wind velocity and density to the HSS at the $200~R_\odot$ boundary. The HSS is continually introduced into the simulation domain through a time-dependent boundary condition which maintains the $200~R_\odot$ boundary at $v_x = 200$~km s$^{-1}$, and a number density of $3$ cm$^{-3}$. These initial profiles of the number density $n = \rho/m$, magnetic field, temperature, and plasma beta ($\beta$) are shown in Fig.~\ref{fig:sim} via the dotted lines.

\subsection{Simulated plasma dynamics}

It is important to note that the simulation is performed in the reference frame of the magnetic obstacle that is co-propagating with the background at a uniform speed. This requires us to specify $v_x = 0$ for the non-HSS plasma. However, the velocity of the HSS is much greater than Alfv\'en speed and the solar wind velocity. This large differential between the solar wind and HSS density and velocity will result in the steepening of velocity and density values which we observe at $t \approx 19.4$ hr via the solid lines in Fig.~\ref{fig:sim}a. As discontinuities in MHD simulations will evolve through a set of wave structures~\citep{Markovskii1996}, the steepening of the incoming HSS into a shock, as seen by the jump in velocity, density, and temperature at the HSS-boundary layer transition, will cause disturbances to be generated at the shock. However, since the HSS and the boundary layer always propagate with super-Alfv\'enic speeds, the HSS will encounter the magnetic obstacle prior to any waves that would be generated at the shock. Thus, the profiles of the magnetic obstacle at $t \approx 19.2$ in Fig.~\ref{fig:sim} will only be a consequence of the HSS interacting with the magnetic obstacle. Note that this setup is employed for a general simulation of the ICME-HSS interaction, and therefore, the absolute values are not crucial.

\subsection{Calculating Dst from simulation data} \label{appsub:dst_sim}
The estimation of the empirical Dst index is discussed in Appendix~\ref{app:dst_pred} through Equations~\ref{eq:Dst-eq}-\ref{eq:dst-difference-eq}. The methodology can be extended to the MHD simulation by considering a virtual observer stationed at $230~R_\odot$. As the simulation is modelled in the frame of reference of the magnetic obstacle, we can extend the simulation results to a solar wind advecting with a constant flow velocity. Thus, by assuming a flow velocity of $v_0 = 450$ km s$^{-1}$, we can utilise Taylor's hypothesis to convert the simulation snapshot at $\approx 19.4$ hr in Fig.~\ref{fig:sim} to time series solar wind data that the virtual observer would encounter at time intervals of $\Delta t = 5$ min. We have verified that the overall trend and estimated values of the Dst remains invariant of changes in the $\Delta t$ parameter. The virtual observer would then encounter the solar wind with expected flow velocities and densities. However, the magnetic fields are an order of magnitude smaller than observed values in the solar wind (Fig.~\ref{fig:overview}). These smaller magnetic field values in the simulation are a consequence of Equation~\ref{eq:pressure-balance} which requires the magnetic energy to be balanced with observationally realistic values of thermal pressure. Therefore, the Dst estimation from simulation data requires scaling the $B_z$ component of the magnetic field in order to ensure it is geoeffective in the $DST^{*1}$ and $DST^{*2}$ forecasting models (see Appendix~\ref{app:dst_pred}) which are empirically fitted for observed solar wind parameters. The unscaled and smaller $B_z$ component results in a minimal contribution of the rate of energy injection in the Dst calculation (Equation~\ref{eq:dst-difference-eq}). Therefore, calculating the Dst index from simulation index requires assuming a solar wind flow velocity $v_0$, and scaling the $B_z$ component of the magnetic field by an order of magnitude.

\section{Supplementary material} \label{app:supp}

\begin{table}
\centering
  \caption{Classification of the ICME-HSS interaction properties used in the calculations shown in Fig.~\ref{fig:corr_matrix} and described in Tab.~\ref{tab:props}, both are listed in Appendix~\ref{app:supp}.}
     \label{tab:list}
     \begin{tabular}{c l}
        \hline
        \noalign{\smallskip}
        Parameter      &  Description  \\
        \noalign{\smallskip}
        \hline
        \noalign{\smallskip}
        $t^{\mathrm{CME}}_{\mathrm{start}}$ & Start time of the magnetic structure (ICME)      \\
        $t^{\mathrm{CME}}_{\mathrm{end}}$ & End time of the ICME      \\
        $t^{\mathrm{HSS}}_{\mathrm{start}}$ & Start time of the undisturbed HSS      \\
        $t_{\mathrm{SymH}}$ & Time of the minimum in the $SymH$ index      \\   
        $SymH$ & Minimum value of the $SymH$ index      \\
        $|B|^{max}_{CME}$ & Maximum of $|B|$ in the ICME     \\
        $|B|^{max}_{BL}$ & Maximum of $|B|$ in the boundary layer     \\
        $B_{z,CME}^{min}$ & Minimum of $B_z$  in the ICME     \\
        $B_{z,BL}^{min}$ & Minimum of $B_z$  in the boundary layer   \\
        $\delta B_{z,BL}$ & Variation of $B_z$ in the boundary layer    \\
        $v^{max}_{HSS}$ & Maximum velocity of the HSS    \\
        $\bar{v}_{CME}$ & Mean velocity of the CME    \\
        $\Delta v_{HSS-CME}$ & $v^{max}_{HSS} - \bar{v}_{CME}$    \\
        $vB^{min}_{z,CME}$ & Minimum of $vB_z$ in the ICME    \\
        $vB^{min}_{z,BL}$ & Minimum of $vB_z$ in the boundary layer    \\
        $n_{max}$ & Maximum density of the interaction    \\
        $P_{max}$ & Maximum dynamic pressure of the interaction    \\
        \noalign{\smallskip}
        \hline
        \multicolumn{2}{l}{\small *Magnetic field data in GSM coordinates} \\
    
     \end{tabular}
\end{table}

\begin{sidewaystable}
\caption{Summary of ICME - HSS interaction events.} 
\label{tab:props} 
\centering
\begin{tabular}{c|c|cccc|ccccccccc}
\hline\hline
N &Year& t$^{\mathrm{CME}}_{\mathrm{start}}$ & t$^{\mathrm{CME}}_{\mathrm{end}}$ & t$^{\mathrm{HSS}}_{\mathrm{start}}$& t$_{SymH}$ & $SymH$ & $\bar{v}_{CME}$ & $v^{max}_{HSS}$ & $|B|^{max}_{CME}$ & $B^{min}_{z,CME}$ & $|B|^{max}_{BL}$ &$\delta B_{z,BL}$  & $n^{max}$ & $P_{max}$ \\
&YYYY&\multicolumn{4}{c|}{TT.MM hh:mm}& [nT] & [km~s$^{-1}$]  & [km~s$^{-1}$] & [nT] & [nT] & [nT] & [nT] & [cm$^{-3}$] & [nPa]    \\ 
\hline
1 & 1999 &16.05 17:10 & 18.05 00:50 & 18.05 08:55 & 18.05 08:00 & -36.0 & 364.7 & 700.2 & 8.6 & -3.3 & 26.7 & 11.0 &60.0 &23.2\ \\ 
2* & 1999 &21.10 04:45 & 22.10 07:05 & 23.10 00:35 & 22.10 07:15 & -218.0 & 464.7 & 697.5 & 37.0 & -31.1 & 19.4 & 2.7 &48.9 &33.9\ \\ 
3 & 1999 &22.11 01:25 & 24.11 06:30 & 24.11 10:45 & 24.11 09:55 & -52.0 & 447.9 & 482.2 & 15.9 & -7.6 & 11.1 & 6.2 &20.3 &9.1\ \\ 
4 & 2000 &24.04 04:40 & 24.04 13:50 & 24.04 23:35 & 24.04 14:00 & -78.0 & 499.5 & 568.4 & 16.8 & -12.7 & 11.0 & 3.2 &8.6 &4.8\ \\ 
5 & 2002 &17.12 21:20 & 19.12 01:10 & 19.12 19:55 & 19.12 18:45 & -79.0 & 375.6 & 551.3 & 18.3 & 1.1 & 23.1 & 7.8 &55.4 &18.5\ \\ 
6 & 2003 &05.08 00:45 & 06.08 02:25 & 06.08 05:40 & 06.08 02:25 & -55.0 & 436.4 & 594.3 & 14.4 & -12.5 & 13.3 & 4.8 &28.1 &11.6\ \\ 
7 & 2005 &10.07 11:45 & 12.07 09:35 & 12.07 16:05 & 12.07 05:30 & -96.0 & 428.1 & 561.6 & 24.9 & -22.2 & 15.5 & 5.1 &30.3 &13.3\ \\ 
8 & 2006 &30.09 09:05 & 30.09 22:50 & 01.10 09:50 & 01.10 03:45 & -57.0 & 394.1 & 588.8 & 18.4 & -8.1 & 14.1 & 3.5 &33.4 &10.4\ \\ 
9 & 2007 &20.11 00:05 & 20.11 12:30 & 20.11 22:10 & 20.11 20:15 & -66.0 & 461.4 & 684.1 & 20.3 & -15.4 & 15.6 & 3.8 &31.4 &11.7\ \\ 
10 & 2009 &20.07 16:40 & 22.07 06:40 & 22.07 11:05 & 22.07 05:55 & -93.0 & 326.6 & 484.6 & 17.6 & -16.9 & 17.2 & 8.1 &46.8 &12.1\ \\ 
11 & 2011 &04.02 13:00 & 04.02 21:35 & 05.02 05:10 & 04.02 21:20 & -67.0 & 418.3 & 663.9 & 23.0 & -19.1 & 16.2 & 3.4 &39.7 &14.7\ \\ 
12 & 2012 &09.07 00:50 & 10.07 02:15 & 10.07 12:15 & 09.07 21:50 & -78.0 & 405.1 & 548.1 & 12.5 & -11.6 & 11.4 & 2.7 &18.3 &5.7\ \\ 
13 & 2012 &13.11 08:55 & 14.11 09:25 & 14.11 21:40 & 14.11 07:25 & -117.0 & 382.9 & 491.0 & 22.9 & -19.1 & 12.2 & 5.5 &26.9 &8.9\ \\ 
14 & 2013 &01.05 12:00 & 01.05 20:15 & 02.05 05:10 & 01.05 19:10 & -67.0 & 426.7 & 506.4 & 11.1 & -9.6 & 6.9 & 2.3 &5.7 &2.2\ \\ 
15 & 2013 &08.11 21:50 & 09.11 07:25 & 09.11 16:00 & 09.11 08:15 & -78.0 & 411.9 & 627.8 & 15.2 & -14.3 & 12.6 & 4.7 &30.1 &10.4\ \\ 
16 & 2015 &03.01 13:05 & 04.01 21:35 & 05.01 01:30 & 04.01 16:30 & -78.0 & 426.1 & 564.1 & 11.4 & -9.8 & 11.5 & 4.7 &14.2 &5.1\ \\ 
17 & 2015 &18.05 19:30 & 19.05 02:30 & 19.05 14:30 & 19.05 02:55 & -64.0 & 455.1 & 544.2 & 18.5 & -15.4 & 15.7 & 3.8 &13.8 &5.8\ \\ 
18 & 2015 &07.08 15:40 & 08.08 14:15 & 09.08 05:30 & 08.08 21:05 & -30.0 & 511.3 & 562.5 & 11.0 & -10.3 & 8.1 & 2.5 &10.0 &6.5\ \\ 
19 & 2015 &16.08 01:25 & 16.08 08:10 & 16.08 20:50 & 16.08 07:30 & -93.0 & 499.7 & 597.1 & 12.8 & -10.6 & 10.6 & 3.4 &9.9 &5.1\ \\ 
20*² & 2015 &27.08 07:20 & 28.08 12:20 & 28.08 23:40 & 28.08 08:40 & -90.0 & 352.0 & 476.9 & 14.2 & -12.2 & 17.4 & 6.7 &39.5 &8.9\ \\ 
21 & 2016 &06.03 05:35 & 06.03 18:05 & 07.03 04:15 & 06.03 21:20 & -109.0 & 376.8 & 610.8 & 20.4 & -17.7 & 20.1 & 5.3 &30.9 &13.1\ \\ 
22 & 2016 &02.08 15:10 & 03.08 05:15 & 03.08 13:00 & 03.08 06:45 & -62.0 & 422.6 & 575.8 & 25.1 & -22.4 & 24.0 & 6.7 &36.0 &12.1\ \\ 
23 & 2016 &10.11 01:25 & 10.11 17:15 & 10.11 22:25 & 10.11 15:20 & -55.0 & 362.6 & 486.3 & 13.7 & -11.6 & 14.1 & 4.1 &39.0 &11.1\ \\ 
\hline
\multicolumn{15}{l}{\small *Excluded from the statistical correlation analysis due to the extremely strong $SymH$ index and $Bz$ value.} \\
\multicolumn{15}{l}{\small *²Complex event with multiple minima in the $SymH$ index.} \\
\end{tabular}
\end{sidewaystable}

\begin{figure}
\centering \includegraphics[width=1\linewidth,angle=0]{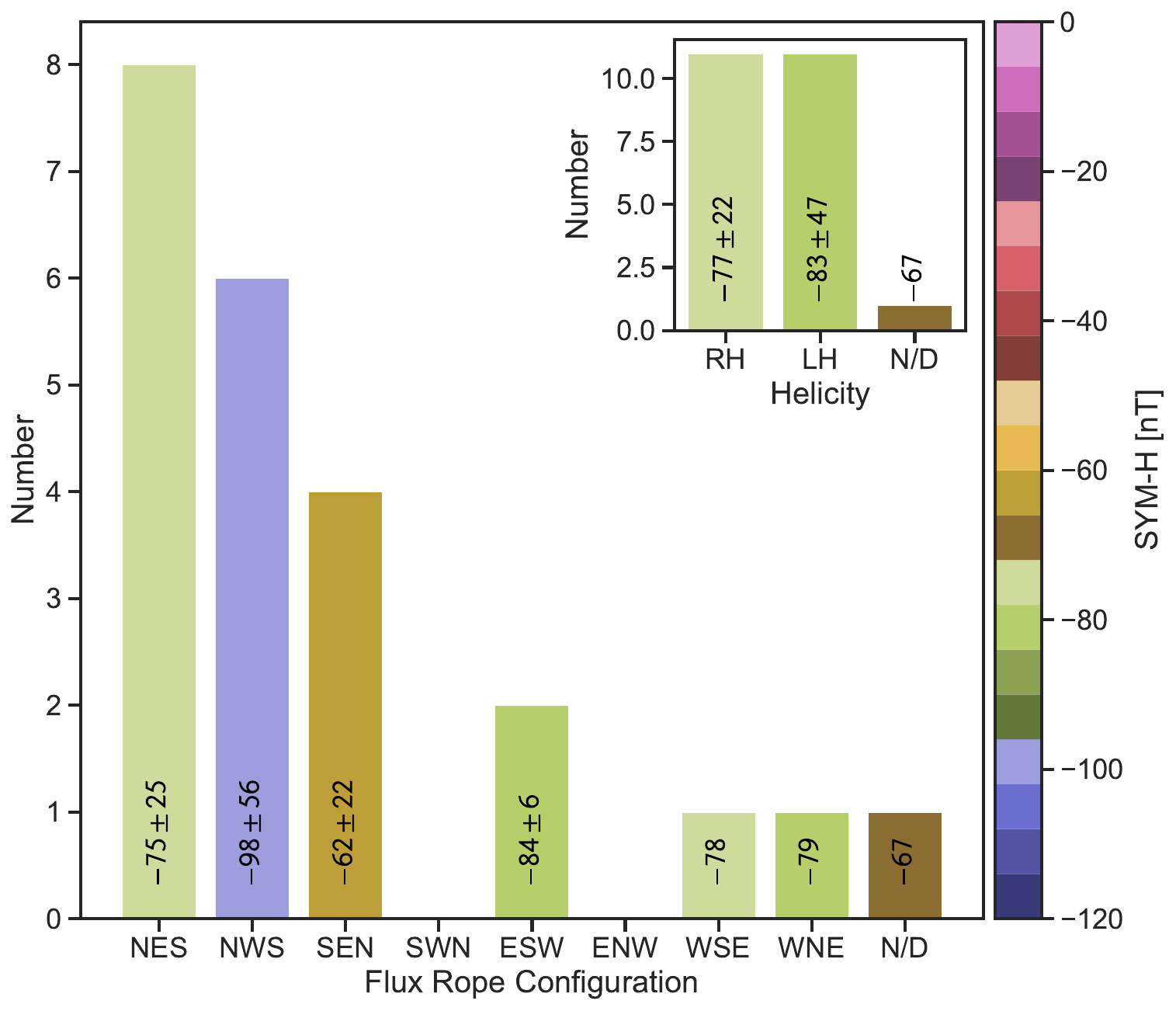}
\caption{Distribution of flux tube orientations. The orientations are labeled according to the convention in \cite{1998bothmer+schwenn}. N, S, E, W correspond to North ($B_z > 0$), South ($B_z < 0$), East ($B_y > 0$), and West ($B_y < 0$). In the abbreviation, the first to third letters represent the variation of the magnetic field vector in the respective component, and the middle letter indicates the direction of the magnetic field on the flux tube axis. The inset shows the distribution of the flux rope helicity between left-handed (LH) and right-handed (RH) flux ropes. }\label{fig:FR_orientation}

\end{figure}

\begin{figure}
\centering \includegraphics[width=1\linewidth,angle=0]{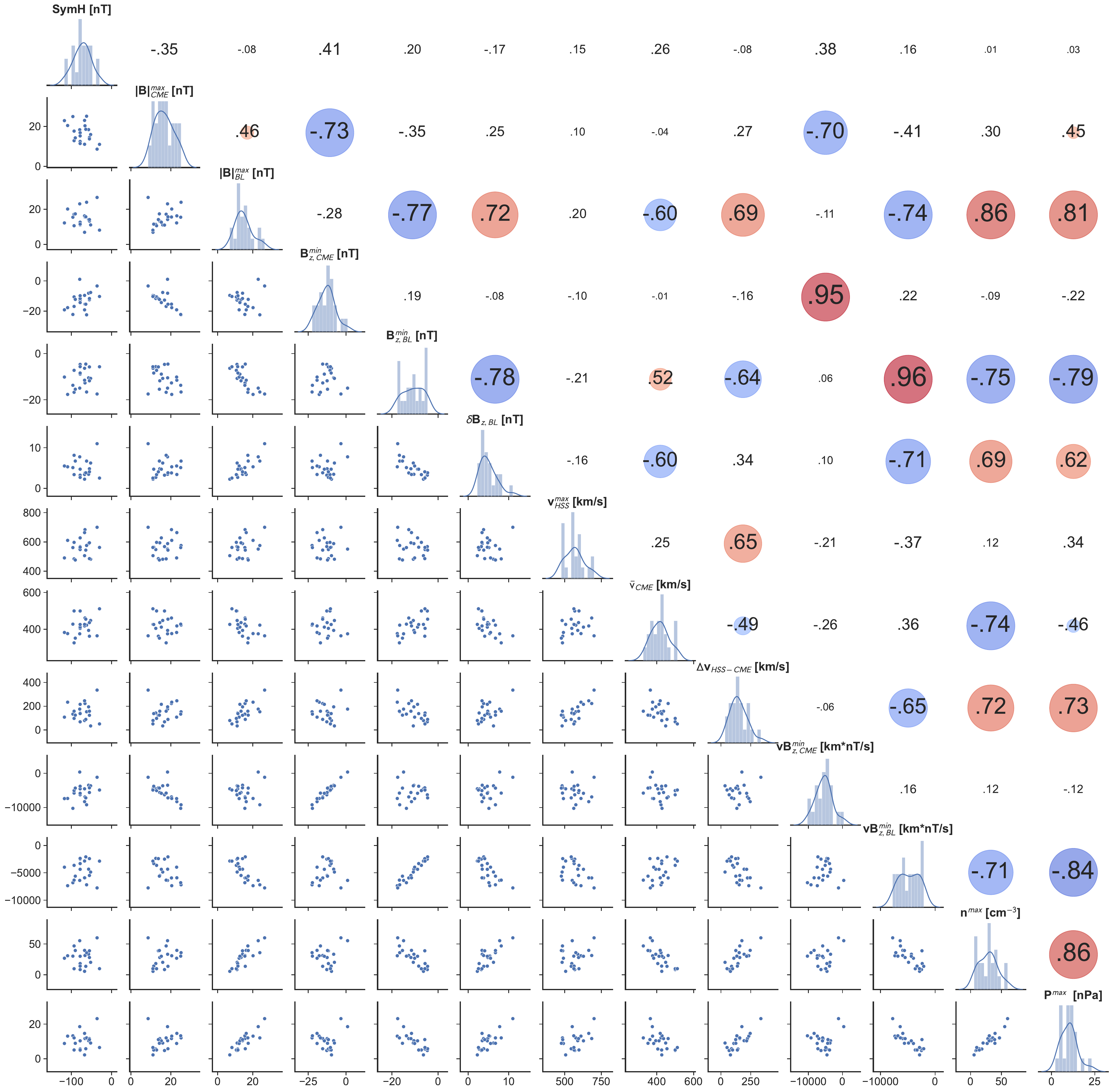}
\caption{Correlation matrix depicting the properties of the HSS-ICME interaction. The Spearman correlation coefficient is presented above the diagonal. The size of the circle indicates the significance of the correlation ($p$-value), with larger circles denoting $p < 0.001$ and absence of circles indicating $p > 0.05$, respectively. Corresponding scatter plots are displayed below the diagonal. The distribution of the parameters is shown along the diagonal.}\label{fig:corr_matrix}
\end{figure}



\bibliography{biblio}{}
\bibliographystyle{aasjournal}



\end{document}